\documentstyle[multicol,prl,aps,graphics]{revtex}

\begin{document}
\draft
\title{Band-structure trend in hole-doped cuprates and correlation with $T_{c\max }$%
.}
\author{E. Pavarini, I. Dasgupta\cite{byline1}, T. Saha-Dasgupta\cite{byline2}, O.
Jepsen, and O.K. Andersen.}
\address{Max-Planck Institut f\"{u}r Festk\"{o}rperforschung,
D-70506 Stuttgart, Germany.}
\date{\today}
\maketitle

\begin{abstract}
By calculation and analysis of the bare conduction bands in a large number
of hole-doped high-temperature superconductors, we have identified the
energy of the so-called axial-orbital as the essential, material-dependent
parameter. It is uniquely related to the range of the intra-layer hopping.
It controls the Cu $4s$-character, influences the perpendicular hopping, and
correlates with the observed $T_{c}$ at optimal doping. We explain its
dependence on chemical composition and structure, and present a generic
tight-binding model.
\end{abstract}


\begin{multicols}{2}

The mechanism of high-temperature superconductivity (HTSC) in the hole-doped
cuprates remains a puzzle\cite{science00}. Many families with CuO$_{2}$%
-layers have been synthesized and all exhibit a phase diagram with $T_{c}$
going through a maximum as a function of doping. The prevailing explanation
is that at low doping, superconductivity is destroyed with rising
temperature by the loss of phase coherence, and at high doping by
pair-breaking\cite{Emery95}. For the {\em materials-}dependence of $T_{c}$
at optimal doping, $T_{c\max },$ the only known, but not understood,
systematics is that for materials with multiple CuO$_{2}$-layers, such as
HgBa$_{2}$Ca$_{n-1}$Cu$_{n}$O$_{2n+2},$ $T_{c\max }$ increases with the
number of layers, $n,$ until $n\sim $3. There is little clue as to why for $%
n $ fixed, $T_{c\max }$ depends strongly on the family, {\it e.g.} why for $n
$=1, $T_{c\max }$ is 40\thinspace K for La$_{2}$CuO$_{4}$ and 85\thinspace K
for Tl$_{2}$Ba$_{2}$CuO$_{6},$ although the Neel temperatures are fairly
similar. A wealth of structural data has been obtained, and correlations
between structure and $T_{c}$ have often been looked for as functions of
doping, pressure, uniaxial strain, and family. However, the large number of
structural and compositional parameters makes it difficult to find what
besides doping controls the superconductivity. Insight was recently provided
by Seo et al.\cite{Locquet00} who grew ultrathin epitaxial La$_{1.9}$Sr$%
_{0.1}$CuO$_{4}$ films with varying degrees of strain and measured all
relevant structural parameters and physical properties. For this
single-layer material it was concluded that the distance between the charge
reservoir and the CuO$_{2}$-plane is the key structural parameter
determining the normal state and superconducting properties.

Most theories of HTSC are based on a Hubbard model with {\em one} Cu$%
\,d_{x^{2}-y^{2}}$-like orbital per CuO$_{2}$ unit. The one-electron part of
this model is, in the ${\bf k}$-representation: 
\begin{eqnarray}
\varepsilon \left( {\bf k}\right) &=&-2t\left( \cos k_{x}+\cos k_{y}\right)
+4t^{\prime }\cos k_{x}\cos k_{y}  \nonumber \\
&&-2t^{\prime \prime }\left( \cos 2k_{x}+\cos 2k_{y}\right) +...\,,
\label{e1}
\end{eqnarray}
with $t,\,t^{\prime },\,t^{\prime \prime },...$ denoting the hopping
integrals $\left( \geq 0\right) $ on the square lattice (Fig.\thinspace \ref
{hop14}). First, only $t$ was taken into account, but the consistent results
of local-density approximation (LDA) band-structure calculations\cite
{ttprime} and angle-resolved photoemission spectroscopy (for overdoped,
stripe-free materials)\cite{ARPES}, have lead to the current usage of
including also $t^{\prime },$ with $t^{\prime }/t\sim $0.1 for La$_{2}$CuO$%
_{4}$ and $t^{\prime }/t\sim $0.3 for YBa$_{2}$Cu$_{3}$O$_{7}$ and Bi$_{2}$Sr%
$_{2}$CaCu$_{2}$O$_{8},$ whereby the constant-energy contours of expression (%
\ref{e1}) become rounded squares oriented in respectively the [11]- and
[10]-directions. It is conceivable that the materials-dependence enters the
Hamiltonian primarily via its one-electron part (\ref{e1}), and that this
dependence is captured by LDA calculations. But it needs to be filtered out:

\begin{figure}
\centerline{\resizebox{1.7in}{!}{{\includegraphics{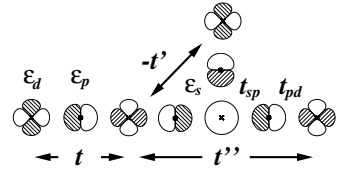}}}\\[2ex]}
\caption[]{\label{hop14}
Relation between the one-orbital model $\left( t,t^{\prime
},t^{\prime \prime },...\right)$ and the nearest-neighbor four-orbital model 
\protect\cite{ttprime} $\left( \protect\varepsilon _{d}-\protect\varepsilon
_{p}\sim 1\,{\rm eV,}\;t_{pd}\sim 1.5\,{\rm eV,}\; \protect\varepsilon _{s}-%
\protect\varepsilon_{p}\sim 16-4\,{\rm eV,}\; t_{sp}\sim 2\,{\rm eV}\right).$
 }            
\end{figure}
The LDA band structure of the best known, and only stoichiometric optimally
doped HTSC, YBa$_{2}$Cu$_{3}$O$_{7},$ is more complicated than what can be
described with the $t$-$t^{\prime }$ model. Nevertheless, careful analysis
has shown\cite{ttprime} that the {\em low}-energy, {\em layer}-related
features, which are the only generic ones, can be described by a {\em nearest%
}-neighbor, tight-binding model with {\em four }orbitals per layer (Fig. \ref
{hop14}), Cu$\,d_{x^{2}-y^{2}},$ O$_{a}$\thinspace $p_{x},$ O$_{b}\,p_{y},$
and Cu\thinspace $s$, with the interlayer hopping $t_{ss}^{\perp }$ 
proceeding via the diffuse
Cu\thinspace $s$-orbital 
whose energy $\varepsilon _{s}$ is several eV above the
conduction band. Also the intralayer hoppings $t^{\prime },\,t^{\prime
\prime },...\,$beyond nearest neighbors in (\ref{e1}) proceed 
via Cu\thinspace $s.$
The constant-energy contours, $\varepsilon _{i}\left( {\bf k}\right) $=$%
\varepsilon ,$ of this model could be expressed simply as\cite{ttprime}: 
\begin{equation}
1-u-d\left( \varepsilon \right) +\left( 1+u\right) p\left( \varepsilon
\right) =\frac{v^{2}}{1-u+s\left( \varepsilon \right) }  \label{e3}
\end{equation}
in terms of the coordinates $u\equiv \frac{1}{2}\left( \cos k_{x}+\cos
k_{y}\right) $ and $v\equiv \frac{1}{2}\left( \cos k_{x}-\cos k_{y}\right) ,$
and the quadratic functions 
\[
d\left( \varepsilon \right) \equiv \frac{\left( \varepsilon -\varepsilon
_{d}\right) \left( \varepsilon -\varepsilon _{p}\right) }{4t_{pd}^{2}}\quad 
{\rm and\quad }s\left( \varepsilon \right) \equiv \frac{\left( \varepsilon
_{s}-\varepsilon \right) \left( \varepsilon -\varepsilon _{p}\right) }{%
4t_{sp}^{2}}
\]
which describe the coupling of O$_{a/b}\,p_{x/y}$ to respectively Cu$%
\,d_{x^{2}-y^{2}}$ and Cu$\,s.$ The term proportional to $p\left(
\varepsilon \right) $ in (\ref{e3}) describes the admixture of O$_{a/b}$%
\thinspace $p_{z}$ orbitals for dimpled layers and actually extends the
four-orbital model to a six-orbital one\cite{ttprime}. For $\varepsilon $
near the middle of the conduction band, $d\left( \varepsilon \right)
,\,s\left( \varepsilon \right) ,$ and $p\left( \varepsilon \right) $ are
positive, and the energy dependence of $d\left( \varepsilon \right) $ may be
linearized ($\dot{d}$%
\mbox{$>$}%
0), while that of $s\left( \varepsilon \right) $ and of $p\left( \varepsilon
\right) $ may be neglected. $p$=0 for flat layers and $p$=$s^{2}/\left(
1+s\right) ^{2}$ for layers dimpled so as to yield
extended saddlepoints. The bilayer bonding and
antibonding subbands have $\varepsilon _{s}$-values split by $\mp
t_{ss}^{\perp }.$ Now, if $\varepsilon _{s}$ were infinitely far above the
conduction band, or $t_{sp}$ vanishingly small, the right-hand side of (\ref
{e3}) would vanish, with the result that the constant-energy contours would
depend only on $u.$ The dispersion of the conduction band near the Fermi
level would thus be that of the one-orbital model (\ref{e1}) with $t$=$%
\left( 1-p\right) /4\dot{d}$ and $t^{\prime }$=$t^{\prime \prime }$=0. For
realistic values of $\varepsilon _{s}$ and $t_{sp},$ the conduction band
attains Cu$\,s$-character proportional to $v^{2},$ thus vanishing along the
nodal direction, $k_{x}$=$k_{y},$ and peaking at $\left( \pi ,0\right) ,$
where it is of order 10 per cent. The repulsion from the Cu$\,s$-band lowers
the energy of the van Hove singularities and turns the constant-energy
contours towards [10]. This same $v^{2}$-dependence pertains to the
interlayer splitting caused by $t_{ss}^{\perp }$ in a multilayer material.
In order to go from (\ref{e3}) to (\ref{e1}), 
\begin{equation}
\frac{1}{1-u+s}=\frac{2r}{1-2ru},\quad {\rm with}\quad r\equiv \frac{1/2}{1+s%
},  \label{e2}
\end{equation}
was expanded in powers of $2ru.$ This provided explicit expressions, such as:%
{\it \ }$t=\left[ 1-p+o\left( r\right) \right] /4\dot{d},$ $t^{\prime }=%
\left[ r+o\left( r\right) \right] /4\dot{d},$ and $t^{\prime \prime }=\frac{1%
}{2}t^{\prime }+o\left( r\right) ,$ for the hopping integrals of the
one-orbital model in terms of the parameters of the four(six)-orbital model
and the expansion energy $\sim \varepsilon _{F}$. Note that all intralayer
hoppings beyond nearest neighbors are expressed in terms of the {\em range-}%
parameter $r.$ Although one may think of $r$ as $t^{\prime }/t,$ this holds
only for flat layers and when $r$%
\mbox{$<$}%
0.2. When $r$%
\mbox{$>$}%
0.2, the series (\ref{e1}) must be carried beyond $t^{\prime \prime }.$
Dimpling is seen not to influence the range of the intralayer hopping, but
to reduce $t$ through admixture of O$_{a/b}\,p_{z}.$ In addition, it reduces 
$t_{pd}.$

Here, we shall generalize this analysis to all known families of HTSC
materials using a new muffin-tin-orbital (MTO) method\cite{LMTO} which
allows us to construct minimal basis sets for the low-energy part of an LDA
band structure with sufficient accuracy that we can extract the materials
dependence. This dependence, we find to be contained solely in $\varepsilon
_{s}$,\ which is now the energy of the {\em axial} orbital, a hybrid between
Cu$\,s$, Cu$\,d_{3z^{2}-1},$ apical-oxygen O$_{c}\,p_{z}$, and farther
orbitals on {\it e.g.} La or Hg. The range, $r,$ of the intralayer hopping
is thus controlled by the structure and chemical composition {\em %
perpendicular} to the CuO$_{2}$-layers. It turns out that the materials with
the larger $r$ (lower $\varepsilon _{s})$ tend to be those with the higher
observed values of $T_{c\max }$. In the materials with the highest $T_{c\max
},$ the axial orbital is almost pure Cu$\,4s.$

\begin{figure}
\centerline{\resizebox{3in}{!}{\rotatebox{-90}{\includegraphics{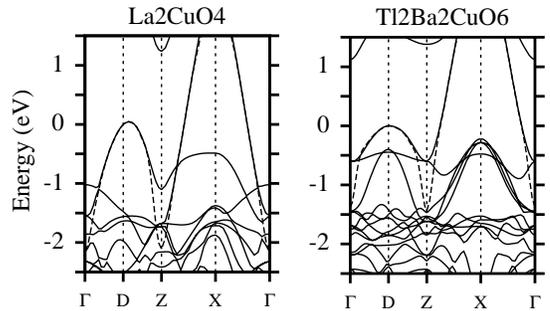}}}\\[2ex]}
\caption[]{\label{LaTlbands}
LDA bands calculated with the NMTO method \protect\cite{LMTO} in
the body-centered tetragonal
structure. The dashed band was obtained using the Bloch sum of MTOs
with N=0 and Cu $d_{x^{2}-y^{2}}$ symmetry at the central site. $\Gamma
\,\left( 0,0,0\right) ,$ ${\rm D}\,\left( \protect\pi ,0,0\right) ,$ ${\rm Z}%
\,\left( 2\protect\pi ,0,0\right) =\left( 0,0,2\protect\pi /c\right) ,$ $%
{\rm X}\,\left( \protect\pi ,\protect\pi ,0\right) .$ 
       }
\end{figure}                     
It should be noted that $r$ describes the {\em shape} of the non-interacting
band in a 1\thinspace eV-range around the Fermi level, whose accurate
position is unknown because we make no assumptions about the remaining terms
of the Hamiltonian, inhomogeneities, stripes, a.s.o.

Fig. \ref{LaTlbands} shows the LDA bands for the single-layer materials La$%
_{2}$CuO$_{4}$ and Tl$_{2}$Ba$_{2}$CuO$_{6}$. Whereas the high-energy band
structures are complicated and very different, the low-energy conduction
bands shown by dashed lines contain the generic features. Most notably, the
dispersion along $\Gamma $DZ is suppressed for Tl$_{2}$Ba$_{2}$CuO$_{6}$
relatively to La$_{2}$CuO$_{4},$ whereas the dispersion along $\Gamma $XZ is
the same. This is the $v^{2}$-effect. The low-energy bands were calculated
variationally with a single Bloch sum of Cu $d_{x^{2}-y^{2}}$-like orbitals,
constructed to be correct at an energy near half-filling. Hence, these bands
agree with the full band structures to linear order and head towards the
pure Cu $d_{x^{2}-y^{2}}$-levels at $\Gamma $ and Z, extrapolating across a
multitude of other bands. This was explained in Ref.\cite{LMTO}. Now, the
hopping integrals $t,$ $t^{\prime },\,t^{\prime \prime },....$ may be
obtained by expanding the low-energy band as a Fourier series. This yields: $%
t$=0.43\thinspace eV in both cases, $t^{\prime }/t$=0.17 for La$_{2}$CuO$%
_{4} $ and 0.33 for Tl$_{2}$Ba$_{2}$CuO$_{6},$ and many further inter- and
intralayer hopping integrals\cite{Indra}.

\begin{figure}
\centerline{\resizebox{2.8in}{!}{\rotatebox{-90}{\includegraphics{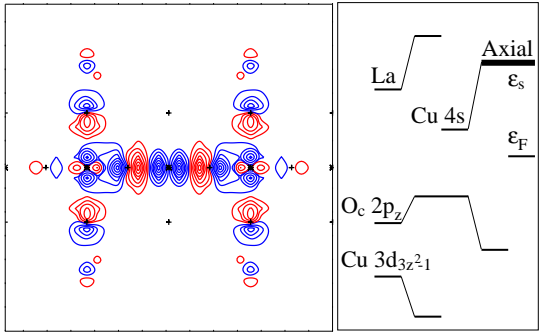}}}\\[2ex]}
\caption[]{\label{xzorbital}
{\it Left:} N=0 MTO describing the Cu $d_{x^{2}-y^{2}}$-like
conduction band in La$_{2}$CuO$_{4}.$ The plane is perpendicular to the
layers and passes through Cu, O$_{a}$, O$_{c},$ and La$. $ {\it Right:}
Schematic diagram giving the energy $\protect\varepsilon _{s}$ of the {\em %
axial} orbital in terms of the energies of its constituents and their
couplings. 
}
\end{figure}

That all these hopping integrals and their materials-dependence can be
described with a generalized four-orbital model, is conceivable from the
appearance of the conduction-band orbital for La$_{2}$CuO$_{4}$ in the $xz$%
-plane (Fig. \ref{xzorbital}). Starting from the central Cu atom and going
in the $x$-direction, we see $3d_{x^{2}-y^{2}}$ antibond to neighboring O$%
_{a}$ $2p_{x},$ which itself bonds to $4s$ and antibonds to $3d_{3z^{2}-1}$
on the next Cu. From here, and in the $z$-direction, we see $4s$ and $%
3d_{3z^{2}-1}$ antibond to O$_{c}$\thinspace $2p_{z},$ which itself bonds to
La orbitals, mostly $5d_{3z^{2}-1}.$ In the $y$-direction, $4s$ antibonds
and $3d_{3z^{2}-1}$ bonds to O$_{b}\,2p_{y}.$ For Tl$_{2}$Ba$_{2}$CuO$_{6},$
we find about the same amount of Cu$\,3d_{x^{2}-y^{2}}$ and O$_{a/b}$%
\thinspace $2p_{x/y}$ character, but more Cu$\,4s,$ negligible Cu$%
\,3d_{3z^{2}-1}$, much less O$_{c}$\thinspace $2p_{z}$, and Tl$\,6s$ instead
of La$\,5d_{3z^{2}-1}$ character. That is, in Tl$_{2}$Ba$_{2}$CuO$_{6}$ the
axial part is mainly Cu$\,4s$.

Calculations with larger basis sets than one MTO per CuO$_{2}$ now confirm
that, in order to localize the orbitals so much that only nearest-neighbor
hoppings are essential, one needs to add {\em one} orbital, Cu axial, to the
three standard ones\cite{Indra}. The corresponding four-orbital Hamiltonian
is therefore the one described above in Fig.\thinspace \ref{hop14} and Eqs. (%
\ref{e3})-(\ref{e2}). Note, that we continue to call the energy of the axial
orbital $\varepsilon _{s},$ and its hopping integral with O$_{a/b}\,p_{x/y}$ 
$t_{sp}.$ Calculations with this basis set for many different materials show
that, of all the parameters, only $\varepsilon _{s}$ varies significantly 
\cite{Indra}. This variation can be understood in terms of the couplings
between the constituents of the axial orbital sketched in the right-hand
panel of Fig. \ref{xzorbital}: We first form the appropriate O$_{c}\,p_{z}$%
-like 5-atom hybrid Cu\thinspace $d_{3z^{2}-1}\,$-\thinspace 2O$%
_{c}\,p_{z}\,\,$-\thinspace 2$\,$La with the energy\cite{Indra} 
\begin{equation}
\varepsilon _{c}=\varepsilon _{\bar{c}}+\left( 1+\frac{t_{sc}}{t_{sp}}\frac{%
t_{pz^{2}}}{t_{cz^{2}}}\right) ^{2}\frac{4\bar{r}t_{cz^{2}}^{2}}{\varepsilon
_{F}-\varepsilon _{z^{2}}}-\frac{t_{c\,La}^{2}}{\varepsilon
_{La}-\varepsilon _{F}},  \label{e7}
\end{equation}
and then couple this to the Cu$\,s$-orbital to yield the energy $\varepsilon
_{s}=\varepsilon _{\bar{s}}+2t_{sc}^{2}/\left( \varepsilon _{F}-\varepsilon
_{c}\right) $ of the axial orbital. Here, the energies of the pure Cu$\,s$-
and O$_{c}\,p_{z}$-orbitals are denoted $\varepsilon _{\bar{s}}$ and $%
\varepsilon _{\bar{c}}$, respectively, while their hopping integral is $%
t_{sc}.$ The energy of the Cu $d_{3z^{2}-1}$-orbital is $\varepsilon
_{z^{2}},$ and its hopping integrals to O$_{a/b}\,p_{x/y}$ and O$_{c}\,p_{z}$
are respectively $t_{pz^{2}}$ and $t_{cz^{2}}.$ In deriving Eqs. (\ref{e3})-(%
\ref{e7}), we have exploited\cite{Indra} that $t_{pz^{2}}^{2}/t_{sp}^{2}\ll 
\frac{\varepsilon _{F}-\varepsilon _{z^{2}}}{\varepsilon _{\bar{s}%
}-\varepsilon _{F}}$ and that $t_{pd}^{2}/t_{sp}^{2}\ll \frac{\varepsilon
_{F}-\left( \varepsilon _{p}+\varepsilon _{d}\right) /2}{\varepsilon
_{F}-\left( \varepsilon _{p}+\varepsilon _{s}\right) /2}.$ Although specific
for La$_{2}$CuO$_{4},$ Eq. (\ref{e7}) is easy to generalize.

\begin{figure}
\centerline{\resizebox{2.1in}{!}{\rotatebox{-90}{\includegraphics{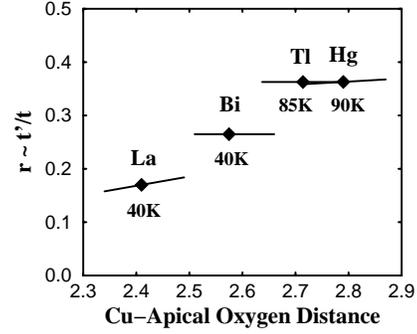}}
}\\ [2ex]} \caption[]
{\label{rvsd}
Calculated range parameter, $r,$ for single-layer materials vs. the
distance (in \AA ) between Cu and O$_{c}$. The lines result from rigid
displacements of O$_{c}$. 
}
\end{figure}    
In Fig. \ref{rvsd} we plot the $r$-values for single-layer materials against
the distance $d_{{\rm Cu-O}_{c}}$ between Cu and apical oxygen. $r$
increases with $d_{{\rm Cu-O}_{c}}$ because $\varepsilon _{s}$ is lowered
towards $\varepsilon _{F}$ when the coupling between O$_{c}\,p_{z}$ and Cu$%
\,d_{3z^{2}-1}/s$ is weakened. Since $t_{cz^{2}}\propto d_{{\rm Cu-O}%
_{c}}^{-4}$ and $t_{sc}\propto d_{{\rm Cu-O}_{c}}^{-2},$ increasing the
distance suppresses the Cu$\,d_{3z^{2}-1}$ content, which is then important
in La$_{2}$CuO$_{4},$ but negligible in Tl$_{2}$Ba$_{2}$CuO$_{6}$ and HgBa$%
_{2}$CuO$_{4}$. This is also reflected in the slopes of the lines in Fig. 
\ref{rvsd} which give $r$ vs. $d_{{\rm Cu-O}_{c}}$ for each material. The
strong slope for La$_{2}$CuO$_{4}$ explains the findings of Seo {\it et al.} 
\cite{Locquet00}, provided that $r$ correlates with superconductivity. That
the Bi-point does not fall on the La-line is an effect of Bi being different
from La: Bi$\,6p_{z}$ couples stronger to O$_{c}\,2p_{z}$ than does
La\thinspace $5d_{3z^{2}-1}.$ The figure shows that upon reaching HgBa$_{2}$%
CuO$_{4}$, $r$ is saturated, $\varepsilon _{s}\sim \varepsilon _{\bar{s}},$
and the axial orbital is almost purely Cu$\,4s.$

\begin{figure}
\centerline{\resizebox{3in}{!}{\rotatebox{-90}{\includegraphics{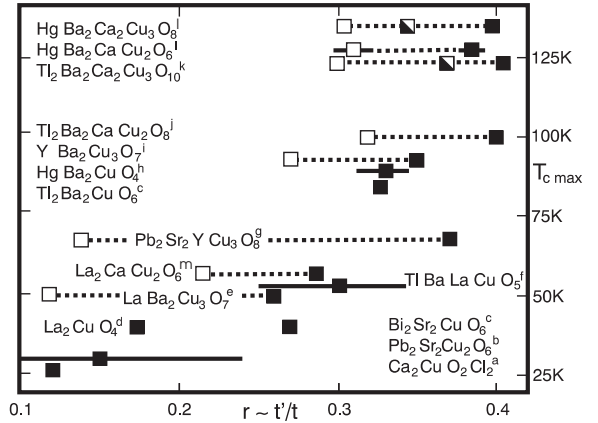}}}\\[2ex]}
\caption[]
{\label{Tcvsr}
Correlation between calculated $r$ and observed $T_{c\max }$. {\it %
Filled squares:} Single-layer materials and most bonding subband for
multilayers. {\it Empty squares:} Most antibonding subband. {\it Half-filled
squares:} Non-bonding subband. {\it Dotted lines} connect subband-values. 
{\it Bars} give $k_{z}$-dispersion of $r$ in primitive tetragonal materials.
a-m \protect\cite
{aComp,bComp,cComp,dComp,eComp,fComp,gComp,hComp,iComp,jComp,kComp,lComp,mComp}
}\end{figure}          
Fig. \ref{rvsd} hints that for single-layer materials $r$ might correlate
with the observed $T_{c\max }$. But the experimental uncertainties of both $%
T_{c\max }$ and the structural parameters are such that we need better
statistics. Therefore, we plot the observed $T_{c\max }$ against the
calculated $r$-values for nearly all known hole-doped HTSCs in 
Fig. \ref{Tcvsr}.
For the single-layer materials, we observe a strong correlation between $r$
and $T_{c\max }$, which seems to be continued in the {\em bonding} subband
for the multilayer materials (filled squares). This indicates that the
electrons are delocalized over the multilayer\cite{Feng}, and that $T_{c\max
}$ increases with the number of layers for the {\em same} reason that it
increases among single-layer materials; the multilayer is simply a means of
lowering $\varepsilon _{s}$ further, through the formation of Cu$\,4s$%
-Cu\thinspace $4s$ bonding states. This is consistent with the celebrated
pressure-enhancement\cite{HgPres} of $T_{c}$ in HgBa$_{2}$Ca$_{2}$Cu$_{3}$O$%
_{8}.$ One might attempt to increase $T_{c\max }$, say for YBa$_{2}$Cu$_{3}$O%
$_{7},$ by substituting Y with a smaller cation, {\it e.g. }Sc. This has not
been done, but a {\em larger} cation, La, was recently inserted\cite{eComp},
and that caused $T_{c\max }$ to drop from 92\thinspace K to 50\thinspace K.
Using the observed structure of LaBa$_{2}$Cu$_{3}$O$_{7}$, we have
calculated the $r$-values and included them in Fig. \ref{Tcvsr}. Here again,
the bonding subband is seen to follow the trend! That $T_{c\max }$
eventually drops for an increasing number of layers, is presumably caused by
loss of phase coherence.

Interlayer coupling in bct La$_{2}$CuO$_{4}$ mainly proceeds 
by hopping from O$%
_{c}\,p_{z}$ at $\left( 0,0,zc\right) $ to its four nearest neighbors at $%
\left( \pm \frac{1}{2},\pm \frac{1}{2},\left( \frac{1}{2}-z\right) c\right),$
and is therefore taken into account by adding to $\varepsilon _{\bar{c}}$
on the right-hand side of (\ref{e7}) the term $-8t_{cc}^{\perp }\,\cos \frac{%
1}{2}k_{x}\cos \frac{1}{2}k_{y}\cos \frac{1}{2}ck_{z}$. In primitive
tetragonal materials, the corresponding term is merely $\propto \cos ck_{z}$
because the CuO$_{2}$-layers are stacked on top of each other, {\it e.g.} in
HgBa$_{2}$CuO$_{4},$ the interlayer coupling proceeds from O$_{c}\,p_{z}$ at 
$\left( 0,0,zc\right) $ via Hg\thinspace $6s/6p_{z}$ at $\left(
0,0,c/2\right) $ to O$_{c}\,p_{z}$ at $\left( 0,0,\left( 1-z\right) c\right)
.$ Periodic interlayer coupling thus makes $\varepsilon _{s}$ depend on $%
k_{z},$ and this passes onto the conduction band a $k_{z}$-dispersion $%
\propto v^{2}\cos \frac{1}{2}k_{x}\cos \frac{1}{2}k_{y}\cos \frac{1}{2}ck_{z}
$ in bct and $\propto v^{2}\cos ck_{z}$ in tetragonal structures. Fig. \ref
{Tcvsr} shows how the $k_{z}$-dispersion of $r$ decreases with contraction
of the axial orbital.

Our identification of an electronic parameter, $r$ or $\varepsilon _{s},$
which correlates with the observed $T_{c\max }$ for all known types of
hole-doped HTSC materials should be a useful guide for materials synthesis
and a key to understanding the mechanism of HTSC. With current ${\bf k}$%
-space renormalization-group methods one could for instance investigate the
effect of the band shape on the leading correlation-driven instabilities\cite
{RG}. Moreover, the possibility that a longer hopping-range leads to better
screening of the Coulomb repulsion, maybe even to overscreening, could be
studied. Increased diagonal hopping, $t^{\prime },$ might lead to higher $%
T_{c\,\max }$ by suppression of static stripe order\cite{Fleck}. The Van
Hove scenario\cite{vH} finds no support in Fig. \ref{Tcvsr} because it is
the saddlepoint of the {\em anti-}bonding band which is at the LDA Fermi
level in YBa$_{2}$Cu$_{3}$O$_{7};$ the bonding band 
is about half-filled and
enhances spin-fluctuations with ${\bf q\approx }\left( \pi ,\pi \right) $%
\cite{Oudovenko}$.$ The propensity to buckling is increased by pushing the
conduction band towards the O$_{a/b}\,p_{z}$-level by lowering
of $\varepsilon _{s}$\cite{ttprime}, 
but recent structural studies\cite{eComp}, as well as
Fig. \ref{Tcvsr}, disprove that static buckling enhances $T_{c\,\max }$,
although dynamical buckling might. The interlayer-pair-tunnelling mechanism
\cite{Chakravarty} is ruled out by the fact that the additional factor $\cos 
\frac{1}{2}k_{x}\cos \frac{1}{2}k_{y}$ attained by $t^{\perp }\left( {\bf k}%
\right) $ in bct materials suppresses the interlayer pair-tunnelling in Tl$%
_{2}$Ba$_{2}$CuO$_{6}$ compared with HgBa$_{2}$CuO$_{4},$ and yet, $T_{c\max
}\sim $90\thinspace K in both cases. That the axial orbital is {\em the}
channel for coupling the layer to its surroundings is supported\cite{Millis}
by the observations that the ${\bf k}$-dependence of the scattering in the
normal state is $v^{2}$-like\cite{ARPES} and that the $c$-axis transport is
strongly suppressed by the opening of a pseudogap\cite{Basov} with similar $%
{\bf k}$-dependence. The axial orbital is also {\em the} non-correlated
vehicle for coupling between oxygens in the layer. Therefore it seems
plausible that contraction of the axial orbital around the CuO$_{2}$-layer,
away from the less perfect doping and insulating layers, 
will strengthen the phase coherence
and thus increase $T_{c\max }.$ Thermal excitation of nodal quasiparticles
\cite{Lee} is, on the other hand, hardly the mechanism by which the
superconducting state is destroyed, because the axial orbital does not
influence the band in the nodal direction. Finally, we note that the
correlation between $r$ and $T_{c\,\max }$ does not extend to electron-doped
cuprates, where the mechanism for superconductivity thus seems to be
different.

Discussions with H. Beck, I. Bozovic, and Z.-X. Shen are gratefully
acknowledged.

\end{multicols}
\end{document}